# Improved Heat and Particle Flux Mitigation in High Core Confinement, Baffled, Alternate Divertor Configurations in the TCV tokamak

Harshita Raj[1], C. Theiler [2], A. Thornton [3], O. Février [2], S. Gorno [2], F. Bagnato [2], P. Blanchard[2], C. Colandrea [2], H. de Oliveira [2], B.P. Duval [2], B. Labit [2], A. Perek [4], H. Reimerdes [2], U. Sheikh[2], M. Vallar [2], B. Vincent [2], the TCV Team [2], the Eurofusion MST1 Team [6]

[1]*Institute for Plasma Research, Gandhinagar, India*
[2]*EPFL, SPC, Lausanne, Switzerland Ecole Polytechnique Fédérale de Lausanne (EPFL), Swiss Plasma Center (SPC), Lausanne, Switzerland*
[3] *CCFE, Culham Science Centre, Abingdon, Oxon, United Kingdom of Great Britain and Northern Ireland*
[4] *DIFFER-Dutch Institute for Fundamental Energy Research, Eindhoven, Netherlands*
[5]*See author list of H. Reimerdes et al. 2022 Nucl. Fusion 62 042018*
[6]*See author list of B. Labit et al 2019 Nucl. Fusion 59 086020*

**Abstract**

Nitrogen seeded detachment has been achieved in the Tokamak a Configuration Variable (TCV) in advanced divertor configurations (ADCs), namely X-divertor and X-point target, with and without baffles in H-mode plasmas with high core confinement. Both ADCs show a remarkable reduction in the inter-ELM particle and heat fluxes to the target compared to the standard divertor configuration. 95-98% of the peak heat flux to the target is mitigated as a synergetic effect of ADCs, baffling, and nitrogen seeded detachment. The effect of divertor geometry and baffles on core-divertor compatibility is investigated in detail. The power balance in these experiments is also investigated to explore the physics behind the observed reduction in heat fluxes in the ADCs.

1. **Introduction**

High particle and heat fluxes deposited in a narrow region at the divertor target challenges the operation of future tokamaks like ITER and DEMO[1,2]. If unmitigated, target heat fluxes would easily exceed material limits of ~$10 MW/m^2$[3]. Detached divertor operation [4–6], where a large fraction of the divertor power is dissipated through impurity radiation, plasma–neutral interactions, and volumetric recombination, is the key to steady-state plasma exhaust in future tokamaks. The detachment process typically sets in at target plasma temperatures below ~5eV [7–9] and results in a reduction of target heat flux, temperature, and ion flux which is important to reduce target erosion [10]. Detachment can be attained by increasing separatrix density [7,12–14] or by impurity seeding [11,15,16]. The latter is predicted to be necessary in a reactor, due to confinement degradation at high upstream density [17] and the Greenwald density limit [18]. It is usually found that divertor detachment degrades the plasma confinement, as the detachment front cools the core plasma through degrading the H-mode pedestal [16,19–22]. To avoid core degradation, ITER plans to operate with a partially detached divertor [23,24], where only the portion of the strike point close to the separatrix is detached [19].

Minimizing any adverse impact on the core confinement while improving the detached operation regime is challenging but crucial in obtaining reliable power exhaust solutions for future reactors. Recent DIII-D experiments, demonstrating fully detached divertor plasmas sustained with high core confinement [25], were a step forward towards this goal. Despite a degraded pedestal associated with divertor detachment, improved core-edge integration was attained by a self-organized plasma state with a low pedestal and a strong ITB, which improved the core confinement [25]. In AUG, real-time control of the X-point radiator with nitrogen seeding facilitated fully detached operation [26] with ~ 95% dissipation of the total power entering the divertor. An ELM-suppressed regime was also observed at a certain height of the radiator above the X-point, with minimal reduction of confinement, combining key requirements for a reactor.

Nevertheless, in DEMO, the exhausted thermal power from the main plasma to the scrape-off layer (SOL) is expected to be 150–300 MW, that is about 1.5–3 times higher than predicted in ITER (~100 MW) [27]. Achieving detached H-mode operation in DEMO will thus be considerably more challenging than for ITER [28,29]. Intense impurity seeding levels are expected to achieve the radiated power fraction of ~95% needed to reach detachment in such conditions, which may be seriously detrimental to the core plasma. In addition, the first wall in DEMO will be considerably more fragile than in ITER due to the tritium breeding structures [30]. A reliable heat exhaust solution for future reactors like DEMO may require alternative divertor concepts that cannot be tested in the conventional divertor used for ITER [31,32]. Exploration and assessment of alternative concepts, in parallel and with reference to the conventional divertor program in ITER, is needed to ensure the timely progress in exhaust solutions for DEMO and future reactors [28,32]. Alternative concepts include Alternative Divertor Configurations (ADCs) employing complex divertor geometries [31,33–35] and/or liquid metal divertor targets [36]. Increasing divertor baffling also offers cooler divertor conditions and promotes access to detachment [37,38]. ADCs aim to reduce the power and particle loads to the target by i) increasing the volume of the radiative region in the SOL and divertor, thereby increasing the dissipated power, ii) extending the wetted area by increased cross-field transport, e.g., through increased turbulence, and iii) enhancing positional stability of the detachment front e.g., by flaring of the flux surfaces in front of the target. Amongst today's most promising ADCs are: the X-Divertor (XD) [39–41] with large poloidal flux expansion ($f_x$) at the outer strike point (OSP), the X-Point Target (XPT) [42] geometry with a secondary X-point near the target, resulting among others, to a large connection length ($L_\parallel$), the Super-X divertor (SXD) with large total flux expansion[43], and the Snowflake divertor (SF) with an additional X-point close to the primary one [44,45]. Unique shaping capabilities of TCV's generous poloidal magnetic field array, and wide divertor diagnostic coverage [46] makes it particularly suitable for exploring the behavior of a range of ADCs with an immediate comparison to a similarly shaped conventional single null (SN) configuration at a proof-of-principle level.

Divertor detachment has been studied in TCV in L-mode confinement, single null configurations [15,47], and for several ADCs: the SF [44,48], XD [34,49], SXD [34], and XPT [34]. An increasingly pronounced roll-over in ion flux at the divertor targets, indicating deeper detachment, and lower peak parallel ion saturation currents as compared to SN were observed in XD configurations. However, the detachment threshold was not strongly affected by $f_x$ [34]. A reduced radiation region location sensitivity to line average density was also observed in the

XD which is crucial for stronger detachment control. The decrease in the detachment threshold with increasing $L_\parallel$ in XPT, as predicted by models, were not observed in TCV [34]. SOLPS-ITER simulations for the SN and XD in DIII-D suggest that the XD achieves an order of magnitude higher carbon emissivity than the SN due to stronger non-coronal effects at smaller field angles [50]. These non-coronal effects are even higher at higher input power leading to lower detachment thresholds in XD, which has been also experimentally observed on DIII-D [41]. Detachment in ELMy and ELM-free H-mode in ADCs were also studied in TCV [51]. A ~3-4 times reduction in peak electron temperature and a ~2 times reduction in the total power load to the outer divertor between ELMs was observed with seeding, whereas the density profile remained broadly unaltered. The impurity emission front comparisons did not reveal any clear increase in divertor cooling in the XD. The back-transition of H-mode to L mode and a shorter H-mode phase caused by strong fueling and seeding were ascribed as the main obstacles in achieving higher divertor heat and particle flux mitigation in H-mode. The recent plasma exhaust (PEX) upgrades at TCV included new in-vessel divertor baffles [38], further diagnostic coverage and increased neutral beam heating capacity [52], improving the access to carry out proof-of-principle experiments of ADCs in H-mode. Models and experimental data from DIII-D indicate that increasing $f_x$ and flaring should work in tandem with divertor closure to achieve a lower detachment threshold [41]. Experiments with baffles in TCV have shown easier access to detachment in L-mode [38] as well as improved core performance at high divertor neutral pressures in H-mode [53]. The exploration and assessment of the combined effect of baffling and ADCs in stable detached ELMy H-mode plasmas is the next logical step in developing advanced exhaust solutions in TCV and is the focus of this paper. Such studies can provide information on configuration(s) with the most promising core-divertor integration for further testing in the Divertor Tokamak Test facility (DTT), which is currently being designed to investigate alternative power exhaust solutions in DEMO-like conditions [54].

In this paper, we present results from $N_2$ seeded detachment experiments achieved in long-legged ADCs, namely the XD and the XPT, and compare them with the SN in stationary ELMy H-mode plasmas with and without divertor baffles for different NBI power levels ($P_{NBI}$). Here, the term ADCs will refer to both the XD and XPT. In particular, a ~2-fold improvement in the inter-ELM heat and particle flux mitigation at the outer divertor with the same amount of $N_2$ seeding in the ADCs in comparison to the SN was achieved, while simultaneously maintaining high core confinement.

The rest of this paper is structured as follows: in section 2 an overview of the experiments and key diagnostics important for this study are described. In section 3, detachment with high core confinement in the baffled ADCs and SN is demonstrated. In section 4, the effect of geometry on target ion flux profiles and integral values in attached and detached phases are presented. In section 5, the effect of geometry on target heat flux profiles is discussed and compared with the effect of baffles. In section 6, the effect of geometry and baffles on the radiated power distribution is presented. In addition, the power balance in the ADCs and the SN is compared with the goal of explaining the physics behind the observed improvement in the exhaust performance of the ADCs. Finally, in section 7, the key results from these experiments are summarised and motivation for future experiments is discussed.

## 2. Experiment overview

These experiments are carried out at the TCV tokamak, which is a carbon walled, medium-sized tokamak (major radius, $R_0$=0.88 m, minor radius, a=0.25 m) with unique flexibility to vary the plasma shape, using 16 independently controlled poloidal field coils [46,52,55]. Figure 1, a)-c), shows the poloidal magnetic geometry of the SN (in red), the XD (in blue), and the XPT (in green) along with radial profiles of their $f_x$ and $L_\parallel$ in d) and e) respectively. In this paper, the term SP2-XPT and SP4-XPT will be used to denote the OSP on the high field side and low field side of the XPT geometry respectively, as marked in figure 1(c). In the XPT, the distance between the two X-points mapped upstream ($dr_{X2}$) is less than 2 mm, i.e., much smaller than the heat flux decay length, $\lambda_q$ (~5 $mm$).

We start from ELMy (Type I) H-mode plasmas similar to those presented in [53] with a plasma current ($I_p$) = 170 $kA$, toroidal magnetic field ($B_0$) = 1.4 $T$, neutral beam heating power ($P_{NBI}$) of 0.75 $MW$ or 1.3 $MW$, $q_{95}$~5 and a Greenwald density fraction $f_G$~0.55. Plasmas in both ADCs and the SN were repeated with and without the polycrystalline graphite baffles recently installed in TCV. The $B_0$ direction is such that the ion grad B-drift is directed downwards, favorable for H-mode access. $I_p$ is in the same direction. These are Deuterium discharges, fuelled from gas valve 1 present at the bottom of the TCV vessel (see fig.1). Detachment is achieved by injecting a prescribed nitrogen gas ($N_2$) pulse from gas valve 3, also situated at the bottom but at the high field side (figure 1). These are piezo-electric gas valves with an integrated pressure sensor, that provides calibrated flow rates into TCV [34]. Density ($n_e$) and temperature ($T_e$) profiles are obtained from Thomson spectroscopy [56] viewing the plasma core and pedestal. The plasma parameters at the divertor targets are measured using an extensive array of wall-embedded Langmuir Probes (LPs) [57,58]. They are operated with a triangular voltage sweep, ranging from -120V to +80 V, at a frequency of 990 Hz. Details on the LP analysis can be found in Ref [57]. The newly installed RADCAM system with five camera modules containing gold foil bolometers with 120 line of sights have been used to estimate the radiated power and its spatial distribution [59]. The line integrated chord intensities thus obtained are tomographically inverted using a minimum Fisher regularization method [60]. MANTIS (Multispectral Advanced Narrowband Tokamak Imaging System) [61] captures images from the divertor region that are inverted using the CalCam package to obtain 2D poloidal maps of the emissivity of the selected radiation lines. In this paper, we will focus on CIII (465.8 nm) line emissivity profiles. A "CIII-front" can be defined as the position where the emissivity along the outer divertor leg decreases by 50% [34,51].

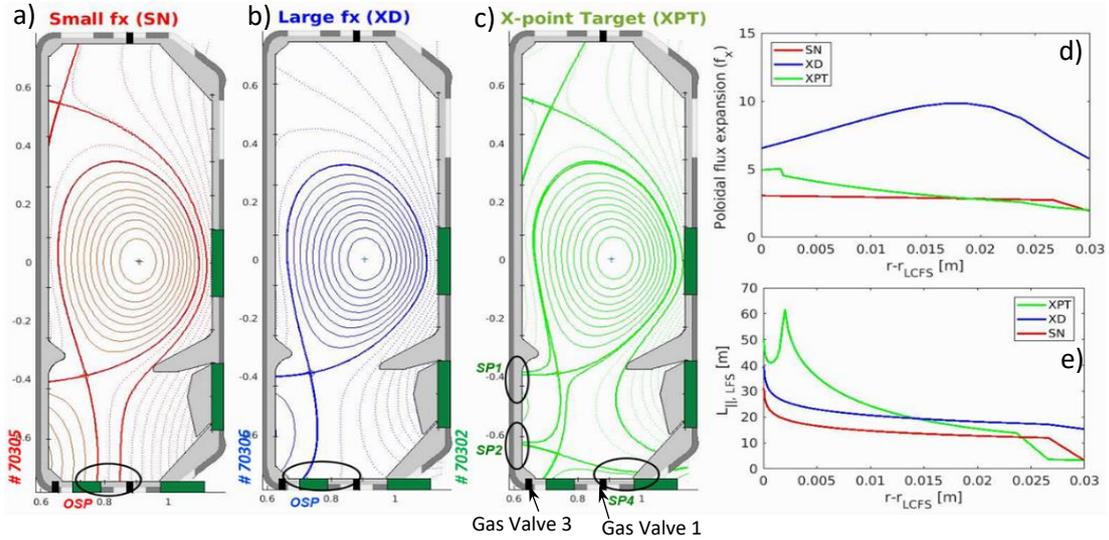

*Figure 1: Plot showing poloidal magnetic geometry of (a) the SN (red), (b) the XD (blue), and (c) the XPT (green) and their radial profile of d) $f_x$ and e) $L_\parallel$ are plotted on the right top and bottom respectively.*

### 3. Divertor detachment in high core confinement, baffled ADCs

Detached ELMy H-mode with high core confinement is achieved with $N_2$ seeding in ADCs and SN geometries with and without baffles in TCV for a range of input powers. The H98 scaling factor is used to assess core performance [62]. It is defined as the ratio of the experimental confinement time ($\tau_E$) and the ITER-98P(y,2) confinement time scaling law ($\tau_E^{IPB98}$), given as

$$\tau_E^{IPB98} = 0.0562\, I^{0.93} B^{0.15} n^{0.41} P^{-0.69} R^{1.97} \kappa^{0.78} \epsilon^{0.58} M^{0.19}$$

where $I\ (MA)$ is the plasma current, $B\ (T)$ is the toroidal magnetic field, $n\ (10^{19}\ m^{-3})$ is the line averaged density, $P\ (MW)$ the input power, $R\ (m)$ the major radius, $\kappa$ the plasma elongation, $\varepsilon$ is the inverse aspect ratio, and $M$ $(\sim m_i/m_p)$ is the hydrogen isotope mass ratio. $\tau_E$ is calculated as $\frac{W_{MHD}}{P_{in}}$, where $P_{in}$ is the input power to the plasma and $W_{MHD}$ the stored plasma thermal energy. $W_{MHD}$ is calculated using electron and ion density and temperature profiles from Thomson spectroscopy and CXRS, respectively. The $W_{MHD}$ thus calculated has also been verified with $W_{MHD}$ obtained from a diamagnetic loop.

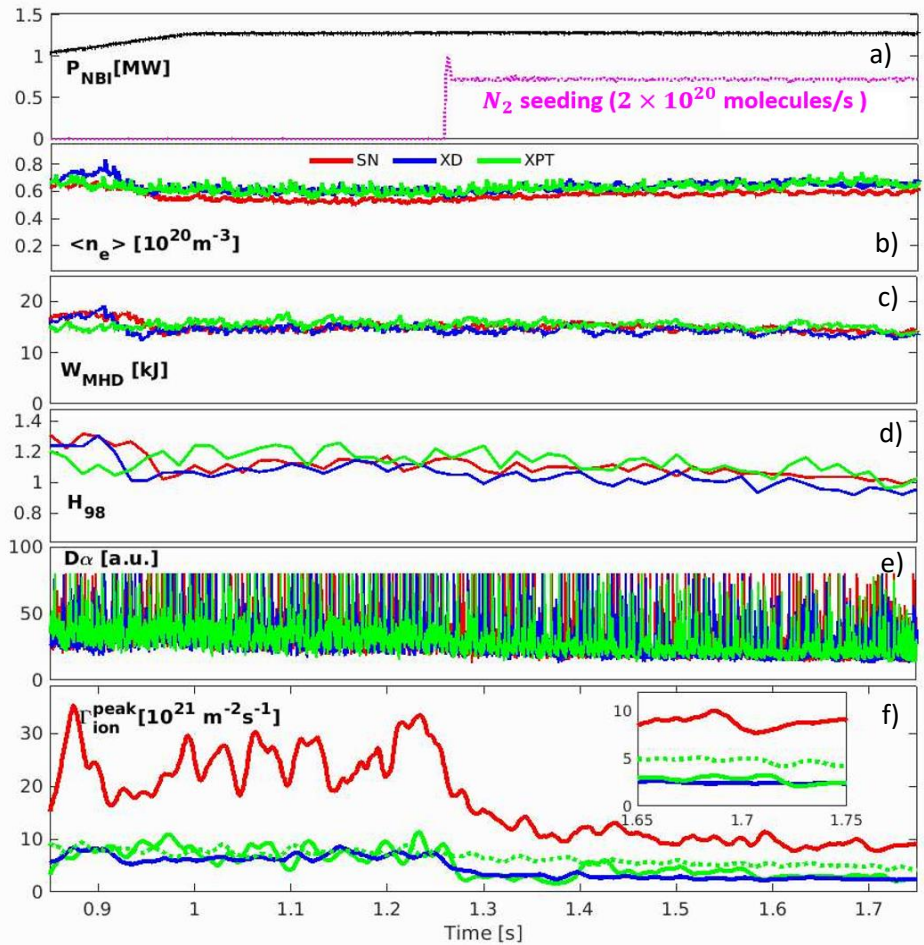

*Figure 2: Plots showing time traces of a) NBI power(black), $N_2$ pulse (magenta), b) line average density, c) stored energy, d) $H_{98}$ factor, e) $D_\alpha$ line emission intensity and f) peak parallel ion flux at OSP for SN (red), XD (blue) and XPT (green).*

Figure 2 shows some major plasma parameters of high power ($P_{NBI} = 1.3\ MW$) baffled shots in ADC and SN geometry with high core confinement ($H_{98} \geq 1$) and stable detached divertor conditions that last for at least 0.4s following the $N_2$ injection. NBI starts at 0.8s, pushing the plasma into H-mode and each geometry stays in a stable ELMy H-mode until the end of the NBI pulse at 1.8s. A constant $N_2$ seeding pulse is applied from 1.25s until the end of discharge. Similar core (line-average) density is maintained in each geometry (fig. 2b) with similar fueling conditions. Before $N_2$ seeding, high core confinement with $H_{98} = 1 - 1.2$ is obtained in each geometry, that decreases slightly following the $N_2$ injection. $H_{98}$ remains $\gtrsim 1$ till the end of the detached H-mode phase, indicating sustained high core confinement (fig 2d). Similar $H_{98}$ factors and $W_{MHD}$ during attached and detached H-mode (fig 2c-d) in both the ADCs and the SN suggests a weak influence of geometry on core confinement. The inter-ELM peak ion flux at the OSP ($\Gamma_{ion,\ OSP}^{peak}$) obtained using LPs (fig 2f) sharply decreases after $N_2$ injection in each geometry, indicating divertor detachment. Both ADCs exhibit 50-70% lower $\Gamma_{ion,\ OSP}^{peak}$ in the attached, as well as detached phase, compared to the SN geometry, while, as noted already, maintaining similar core confinement. During the attached phase, $\Gamma_{ion,\ OSP}^{peak}$ in the XD and both active OSPs of the XPT (SP2-XPT, SP4-XPT) appear quite similar. The effect of $N_2$ seeding on the peak flux at SP4-XPT is weaker as compared to SP2-XPT and the XD, that exhibit the lowest particle fluxes in the detached phase. However, despite weaker detachment at SP4-XPT, its $\Gamma_{ion}^{peak}$ at is half of that in the SN. Detailed discussion on the effect of geometry and baffles on particle and heat fluxes at the target will be presented in section 4.

### *3.1 Effect of geometry on core/pedestal*

The effect of divertor geometry on the pedestal profiles was probed by comparing the pedestal $n_e$, $T_e$ and $P_e$ profiles obtained from TS as presented in fig 3(a-c) in the attached and 3(d-f) in the detached phase. These profiles are obtained from the [0.3-0.9] fraction of the ELM cycle. Normalized poloidal flux, $\rho_\psi = \sqrt{\frac{\psi - \psi_0}{\psi_{x1} - \psi_0}}$, (where $\psi$ is the poloidal flux and $\psi_0$ and $\psi_{x1}$ its values at the magnetic axis and primary X-point, respectively) is used as the radial coordinate. Although the H98 factor and $W_{MHD}$ seem largely unaffected by the divertor geometry, the pedestal $n_e, T_e$ and $P_e$ profiles obtained during inter-ELM time windows show marginally higher $n_{e_{ped}}$, $T_{e_{ped}}$ and $P_{e_{ped}}$ in the ADCs, especially for the XPT compared to the SN in both the attached and detached phases.

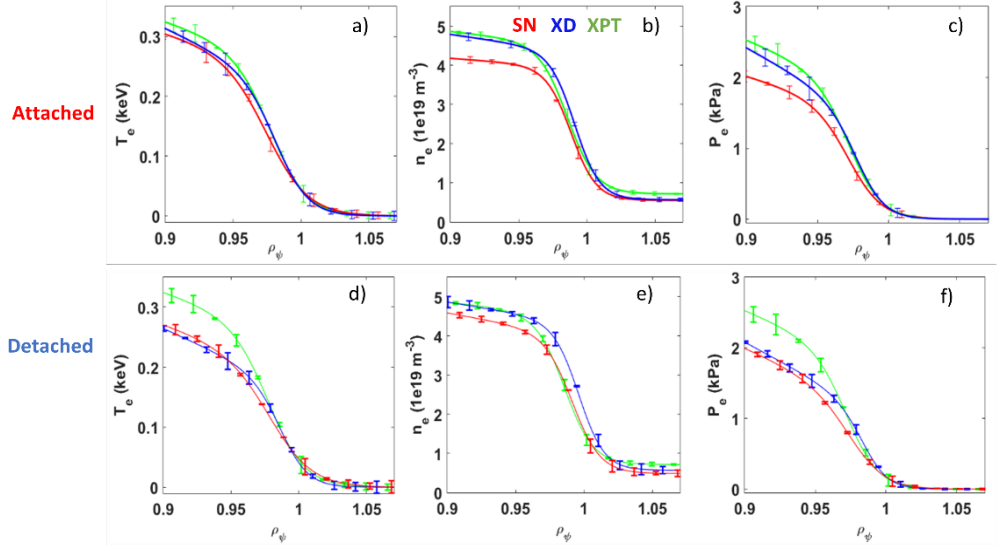

*Figure 3: Plots showing radial inter-ELM pedestal profiles of $T_e$, $n_e$ and $P_e$ in (a-c) attached and (d-f) detached phase, respectively.*

A minor degradation in the $n_{e_{ped}}$ and $T_{e_{ped}}$ is observed in the detached phase in all geometries, which is consistent with the H98 and $W_{MHD}$ trends discussed earlier. It is worth mentioning that although the H98 and $W_{MHD}$ slightly decrease following N2 seeding, the core density increases in all three geometries, as shown in figure 4a.

The effect of geometry on core confinement and pedestal parameters in ADCs was further examined at lower power, $P_{NBI} = 0.75\ MW - 1\ MW$, here again, with and without baffles,

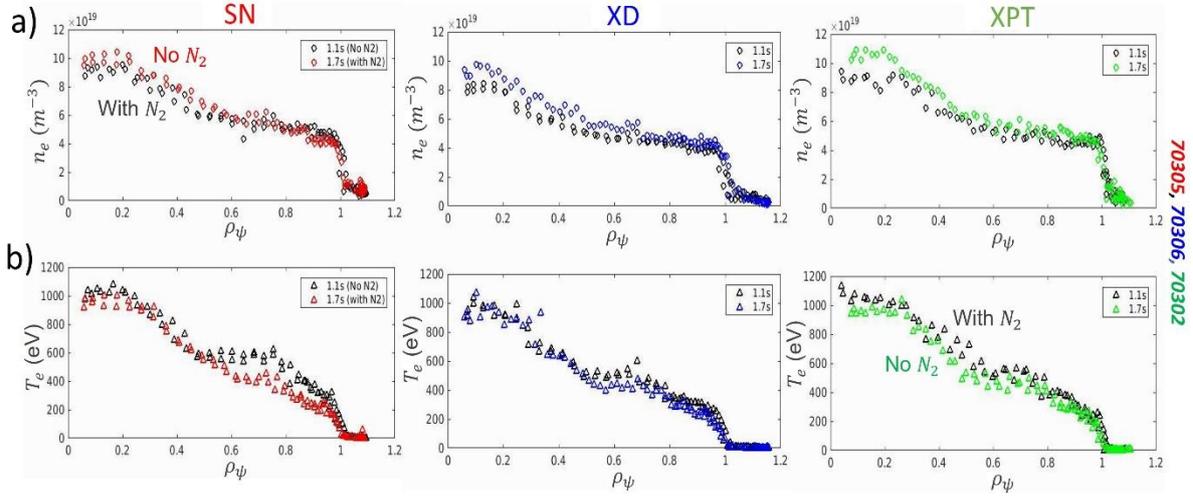

*Figure 4: Plots showing radial profiles of a) $n_e$ and b) $T_e$ in SN, XD and XPT respectively. Black points show the attached, and coloured points show the detached phase. (Same shots as in fig.3)*

and showed similarly small dependencies. The ELM behaviour is another important aspect of H-mode plasmas, as it plays a significant role in the power balance. As we report upon the inter-ELM power and particle exhaust in the following sections, we shall also report on the ELM behaviour changes with the divertor geometry. Figure 5 shows the ELM frequency ($f_{ELM}$), energy lost per ELM ($dW_{ELM}$) and ELM power loss ($P_{ELM}$) for the 1.3 MW baffled shots presented in figure 1 and 2. $dW_{ELM}$ is estimated as the difference in the stored energy, from a diamagnetic loop before and after an ELM following correction of any over-shoot in that measurement [51].

$P_{ELM}$ is estimated as the product of the energy loss per ELM and the ELM frequency ($P_{ELM} = f_{ELM}.dW_{ELM}$). As indicated by figure 5, in the attached phase, $f_{ELM}$ and $P_{ELM}$ is not strongly affected by the divertor geometry, although $f_{ELM}$ is somewhat lower and $dW_{ELM}$ higher in the XPT. From the start of the seeding phase, $f_{ELM}$ increases and $dW_{ELM}$ decreases. By the end of the seeded phase, $dW_{ELM}$ has decreased by 35-40% in all geometries. Despite some variability in $f_{ELM}$ and $dW_{ELM}$, $P_{ELM}$ is similar in all the geometries, dropping by ~30% due to seeding. Unlike AUG experiments [63], where an ELM-suppressed regime is observed at a certain height (~10 cm) of the XPR above the X-point, discharges reported in this paper remain in an ELMy regime in all geometries. The decreasing $dW_{ELM}$ and $P_{ELM}$ and increasing ELM frequency indicate, however, a tendency towards a small ELM regime late in the detached H-mode phase.

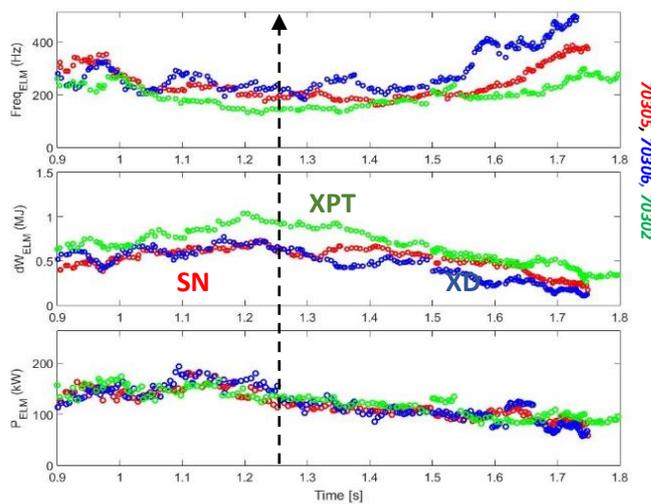

Figure 5: Plots showing the time evolution of the 20 points moving average of a) $f_{ELM}$, b) $dW_{ELM}$ and c)$P_{ELM}$ in SN (red), XD (blue) and c) XPT (green) in 1.3 MW plasmas with constant N2 seeding starting at 1.25s, indicated by dotted black arrow.

### 3.2 Detachment in all geometries verified with MANTIS

Earlier detachment studies [15,34,51,64] showed that the CIII front is a good proxy for the position of the boundary of the cold radiative region along the divertor leg. The displacement of this CIII-front from the target to the X-Point is indicative of divertor cooling, and, thereby, a marker for detachment. Figure 6 (a) shows an inter-ELM MANTIS time frame captured during the attached phase in the SN geometry, where the CIII emission extends along the entire outer divertor leg in contrast to the detached phase, where the CIII front is very close to the X-point in between ELMs (see fig 6 b). Figure 6(c) shows the temporal evolution of the inter-ELM poloidal distance of the CIII front to the primary X-point in 1.3 MW baffled ADCs and SN. A 25 ms median filter was used to obtain the inter-ELM CIII front of each geometry. During the attached phase, the CIII front sits closer to the primary X-point in the ADCs, compared to the SN, indicating colder divertor conditions in the ADCs. As the $N_2$ seeding starts, the CIII front migrates further towards the X-point in each geometry. These results from MANTIS confirm detachment is obtained for the ADCs and the SN. An advantage of ADCs in terms of divertor cooling is also apparent in a facilitated detachment onset. It should be noted that the poloidal leg length of the outer divertor, it should be noted, is different in each geometry. Therefore, the CIII front poloidal distance normalized w.r.t the total poloidal leg length ($CIII\ front_{NM}$) may be a more adequate parameter to compare. This is plotted in figure 6d for the ADCs and the SN in 1.3 MW baffled scenarios with a $N_2$ seeding ramp. In the attached phase, the $CIII\ front_{NM}$ is closest to the X-point in the XD, followed by the XPT and lastly, the SN geometry where the $CIII\ front_{NM}$ is close to the target. This indicates that between the two ADCs, the XD has a colder divertor in the attached phase.

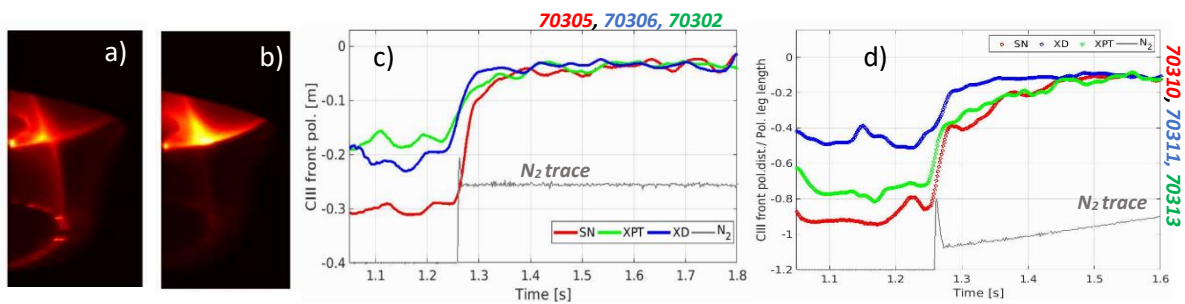

Figure 6: Inter-ELM MANTIS time frame from the SN captured during a) the attached phase and b) detached phase. Plot showing temporal evolution of the inter-ELM (c) poloidal distance of CIII front to the primary X-point and d) CIII front poloidal distance normalized w.r.t the total poloidal leg length in 1.3 MW baffled ADCs and SN.

Together with the LP data (fig.2), these results provide further evidence of the achievement of a stable $N_2$ seeded detached H-mode with high core confinement in the ADCs and the SN. We also presented indications of colder divertor conditions (from CIII front measurements) and lower target peak ion fluxes (fig 2f) in the ADCs. In the following sections, we further examine the LP data, quantifying the advantages of the ADCs over the SN in terms of the particle and heat fluxes on the divertor target.

## 4. Effect of geometry on target ion flux profiles

For a reliable interpretation of LP data, the inter-ELM data in swept mode is carefully extracted for each discharge, and then analysed to obtain target plasma parameters. The details on LPs in TCV and its data analysis are discussed in Ref [59]. Figure 7 shows the radial profiles of the inter-ELM ion flux parallel to the magnetic field ($J_{sat}$) measured at the OSP as a function of $\rho_\psi$ for the SN (fig 7a) and the ADCs (fig.7 b, c, d) during the attached (black) and the detached phases (colored). In the attached H-mode phase, the XD, SP2-XPT and SP4-XPT exhibit substantially (<50%) lower peak $J_{sat}^{peak}$ at the OSP in comparison with the SN. The $J_{sat}^{peak}$ in the XD and SP2-XPT are similar in magnitude, whereas the $J_{sat}^{peak}$ at the SP4-XPT is substantially lower. This advantage of ADC over the SN in terms of lower peak ion fluxes also persists into the detached phase. The effect of seeding at SP4-XPT appears weaker than for the SP2-XPT.

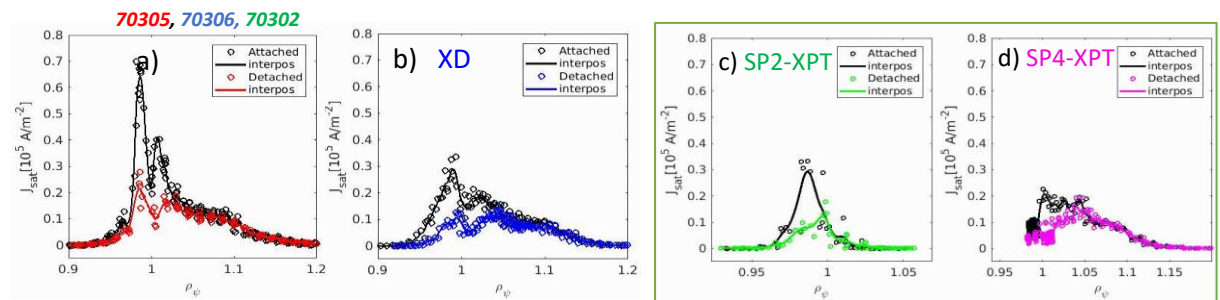

*Figure 7: Plots showing the radial profiles of the inter-ELM ion current parallel to the magnetic field ($J_{sat}$) at the OSP vs. $\rho_\psi$ for a) SN, b) XD, c) SP2-XPT and, d) SP4-XPT during the attached (black) and the detached phase (colored).*

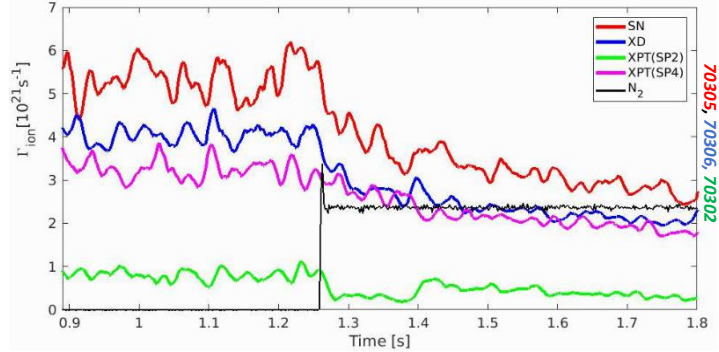

*Figure 8: Plot showing time evolution of the integrated ion flux to the OSP in the SN (red), XD (Blue), SP2-XPT (green) and SP4-XPT (magenta)*

This could be due to the proximity of SP2-XPT to the N2 seeding gas valve (valve3) that is poloidally farther for SP4-XPT. To elucidate any issue of poloidal asymmetry in detachment in the XPT, $N_2$ seeding experiments with multiple poloidal gas valves are planned for future TCV experiments. They are expected to add further important details to the results presented here. Figure 8 show the integrated ion flux ($\Gamma_{ion}$) at the OSP in the SN and the ADCs. The total $\Gamma_{ion}$ in the ADCs is lower than for the SN in both attached and detached phases. In all geometries, $\Gamma_{ion}$ gradually decreases with $N_2$ seeding in the detached phase. It is worth noting that although $\Gamma_{ion}$ at SP2-XPT is much lower than SP4-XPT, the $J_{sat}^{peak}$ at SP2-XPT is similar due to the more peaked profile. We have shown that both ADCs demonstrate a clear advantage over the SN with lower $J_{sat}^{peak}$ (fig 2f and 7) as well as lower $\Gamma_{ion}$ (fig 8) at the OSP in H-mode. However, for attached XPT plasmas, these desirably low $J_{sat}^{peak}$ at both SP2 and SP4 are limited to scenarios with extremely small $dr_{X2}$ (~1 mm), i.e. ~20% of the (upstream) heat flux decay length $\lambda_q$ (~5 $mm$), measured using IR in similar SN geometry. In high $dr_{X2}$ scenarios, $J_{sat}^{peak}$ SP2 is almost the same as in the SN, denoted by grey points in fig 9d and 9e. Again, a similar $J_{sat}^{peak}$ distribution at SP2-XPT and SP4-XPT was obtained by carefully optimizing the distance between the X-points. Thus, as one might expect, the $J_{sat}^{peak}$ ratio at SP2 and SP4 in the XPT geometry strongly depends on the distance between the X-points. Figure 9 (a) shows the distance between the X-points mapped upstream ($dr_{X2}$) for two XPT discharges with similar plasma parameters but different $dr_{X2}$. Figure 9 (b) and 9 (c) show the corresponding geometries. In figure 9 (d) and (e) we show the radial $J_{sat}$ profile for SP2-XPT (magenta), SP4-XPT (blue) and compare them to the $J_{sat}$ in the SN (grey) of the XPT discharges. With high $dr_{X2}$ (5-7 mm), $J_{sat}^{peak}$ at SP2 is as high as in the SN and $J_{sat}^{peak}$ at SP4 is much lower. For smaller $dr_{X2}$ (1-2 mm), $J_{sat}^{peak}$ at SP2 and SP4 are comparable and considerably lower than for SN. From lower parallel particle fluxes in the ADCs reported here, we can also expect lower parallel heat fluxes, discussed in the next section.

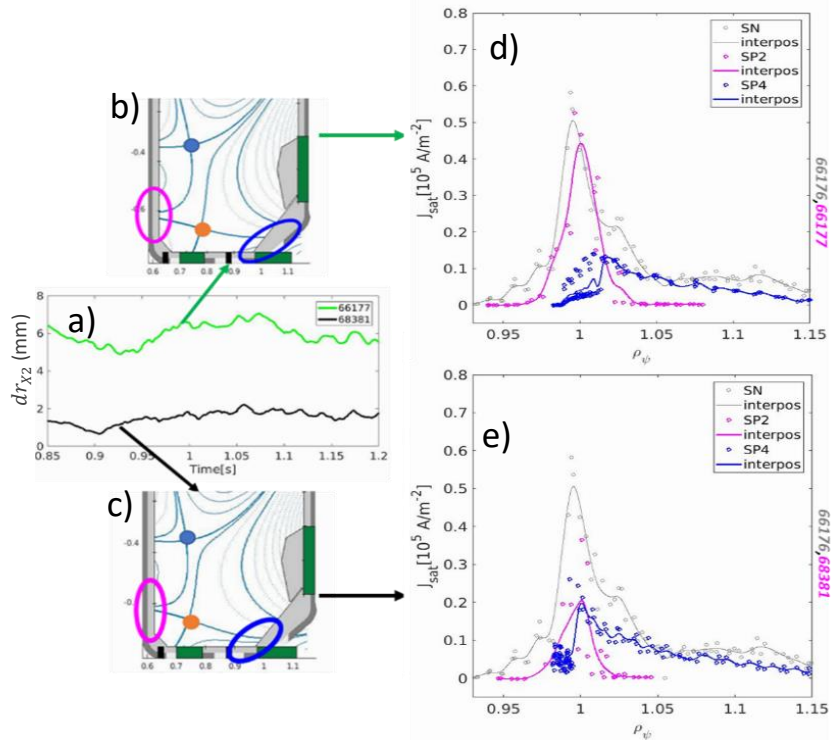

*Figure 9: (a) Time evolution of $dr_{X2}$ for two XPT discharges with similar plasma parameters with high $dr_{X2}$ (green) and low $dr_{X2}$ (black). 2D equilibrium showing both X-points for XPT with (b) high $dr_{X2}$ and (c) low $dr_{X2}$. Radial $J_{sat}$ profile of SP2-XPT (magenta), and SP4-XPT (blue) in (d) high $dr_{X2}$ and (e) low $dr_{X2}$ case, in comparison to the $J_{sat}$ profile in the SN (grey).*

## 5. Improved heat flux mitigation in baffled ADCs
### 5.1 Effect of ADCs & baffles on target heat flux

In supplement to the CIII front measurements indicating colder divertor conditions in the ADCs (section 3.2) and lower peak parallel ion fluxes (section 4), we present more detailed and qualitative evidence from the LPs showing improved heat flux mitigation in ADCs. The inter-ELM parallel heat fluxes at the OSP ($q_{\parallel,OSP}$) is evaluated as $q_{\parallel,OSP} = J_{sat}(\gamma T_e + E_{pot})$, where $E_{pot}$ is the potential energy carried by the incident ions ($E_{pot}$ = 13.6 + 2.2 = 15.8 eV, following Ref. [59]) and, $\gamma$ is the sheath heat transmission coefficient, taken to be $\gamma = 5$. Experiments were performed to compare the impact of divertor geometry (XD and XPT) and baffles on peak heat flux mitigation. We will first discuss the impact of ADCs and baffles in the attached phase. Figure 10 shows the radial profiles of $q_{\parallel,OSP}$ for the ADCs and the SN without baffles (top row) and with baffles (bottom row). The $q_{\parallel,OSP}$ profiles plotted here are obtained from the attached H-mode phase in low power ($P_{NBI}$= 0.75 MW) plasmas. We note that here, $q_{\parallel,OSP}$ is shown rather than the perpendicular heat flux to the wall, to avoid effects related to varying the magnetic field incidence angles at the target.

As expected from the $J_{sat}$ profiles, the peak $q_{\parallel,OSP}$ ($q_{\parallel,OSP}^{peak}$) is highest in the unbaffled SN with $q_{\parallel,OSP}^{peak} \sim 4\ MW/m^2$. When increasing $f_x$ (in unbaffled XD), $q_{\parallel,OSP}^{peak}$ is reduced by ≈60% reaching $1.6\ MW/m^2$. Conversely, with baffles (in baffled SN) one obtains ≈40% lower $q_{\parallel,OSP}^{peak}$, $2.5\ MW/m^2$. As a cumulative impact of baffled and increased $f_x$, ≈85% $q_{\parallel,OSP}^{peak}$ mitigation was achieved, with $q_{\|,OSP}^{peak} \sim 0.6\ MW/m^2$. By adding a secondary X-point (unbaffled XPT), we could obtain ≈75% lower $q_{\|,OSP}^{peak}$ at both SP2 and SP4, provided that $dr_{X2}$ ≤1 mm. $q_{\|,OSP}^{peak}$ at SP2-XPT with high $dr_{X2} (\geq 2$ mm) was found to be up to two times higher, but still lower than the SN despite similar $J_{sat}$ (see fig. 8). For this XPT scenario, no baffled case is available with closely matched $dr_{X2}$ and core density.

These findings reveal clear reduction in heat fluxes at the target individually by the ADCs and the baffles compared to the unbaffled SN. They also indicate that the impact of ADCs is stronger here than baffling in terms of the target heat flux reduction. When these two means are combined in the baffled XD, it leads to remarkably low $q_{\|,OSP}^{peak}$.

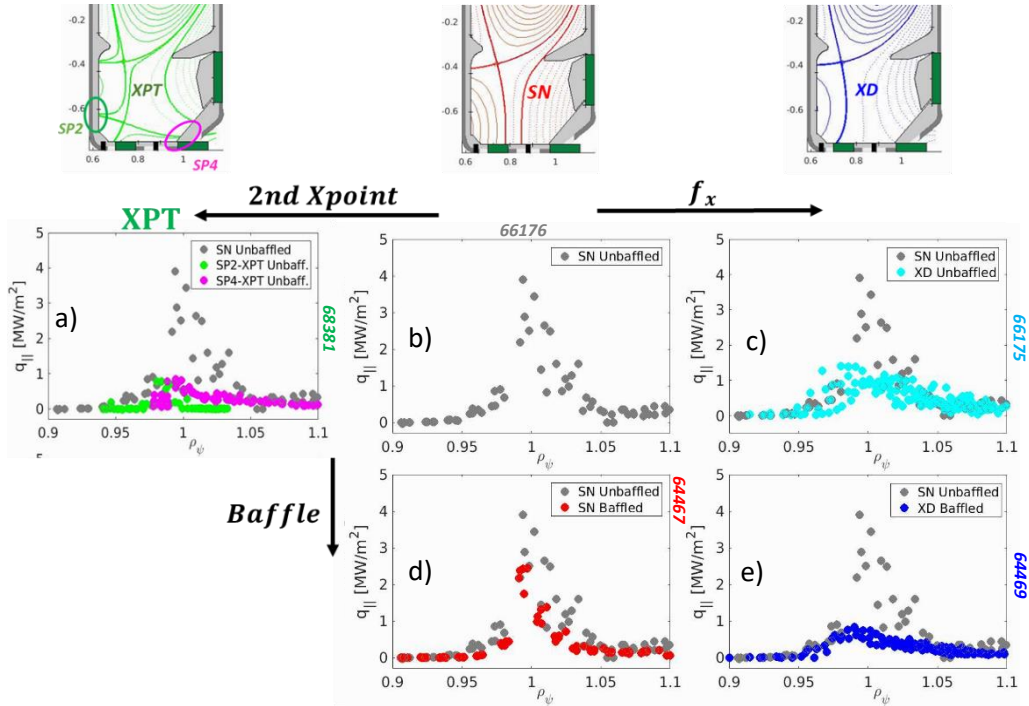

*Figure 10: Plots showing radial profile of $q_{\|,OSP}$ in a) Unbaffled XPT, b) Unbaffled SN, c) Unbaffled XD, d) Baffled SN and e) Baffled XD in $P_{NBI}$= 0.75MW shots.*

Although at these low powers ($P_{NBI}$= 0.75 MW), the ADCs and the SN plasmas exhibit clear signs of detachment after N2 seeding i.e., a sharp reduction in $J_{sat}$ and swift migration of the CIII front to the X-point, it was difficult to assess the effect of geometry on heat flux mitigation in the detached H-mode due to very low $q_{\|,OSP}^{peak}$ ($< 0.4\ MW/m^2$) in detached baffled ADCs, that limits LP data interpretation. In the following section, we look at target heat fluxes in the high-power ($P_{NBI}$= 1.3 MW) for both attached and detached ADCs.

### 5.2 Detachment in high power, baffled, H-mode ADCs

$q_{\|,OSP}^{peak}$ is, as expected, higher in the high-power baffled SN plasmas (~5 $MW/m^2$) than the lower power baffled SN (~3 $MW/m^2$). This facilitates a reliable comparison of the detached phases. Figure 11 plots the radial profiles of $q_{\|,OSP}$ for the SN and the ADCs with baffles in the attached (black) and detached (color) phases.

When attached, the $q_{\parallel,OSP}^{peak}$ in the XD remains ~50% lower than in SN, whereas in the XPT, $q_{\parallel,OSP}^{peak}$ is ~65% and ~80% lower at the SP2 and SP4, respectively. This agrees with our findings at lower power. The difference in $q_{\parallel,OSP}^{peak}$ between SP2 and SP4 may be attributed to a slightly higher $dr_{X2}$(2-3 mm) compared to $dr_{X2}$(1-2 mm) in low-power XPT (fig 10a), where SP2 and SP4 show similar peak heat fluxes.

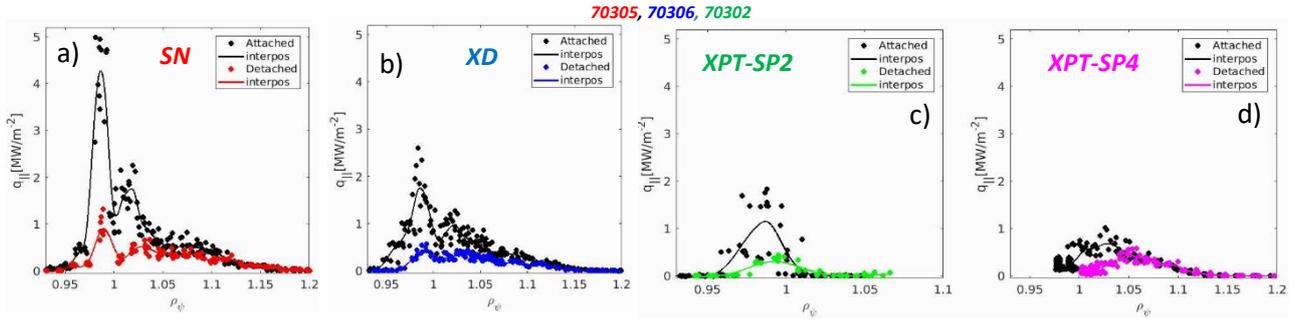

Figure 11: Plots showing radial profile of $q_{\parallel,OSP}$ in baffled a) SN, b) XD c) SP2-XPT and d) SP4-XPT in $P_{NBI}$= 1.3 MW shots.

Next, we consider the heat flux mitigation with ADCs in the detached phase. Coloured profiles in fig. 11 are taken from a stable detached time window (~1.6s) when the $J_{sat}$ is at its minimum, the CIII front is at the X-point, and $H_{98}$~1. An almost 80% reduction in $q_{\parallel,OSP}^{peak}$ is seen with detachment in the SN and the XD. Maximum reduction in the $q_{\parallel,OSP}$ occurs close to the strike point for $0.95 < \rho_\psi < 1.05$. The resulting $q_{\parallel,OSP}^{peak}$ in the XD is ~60% lower than the SN in the detached phase. In the XPT, $q_{\parallel,OSP}^{peak}$ is reduced by ~70% at SP2, but only by ~40% at SP4. Nevertheless, $q_{\parallel,OSP}^{peak}$ at both SP2 and SP4 remains considerably lower (~50%) than for the detached SN. Also, despite higher $dr_{Xpt}$(~3 mm) in detached phase, $q_{\parallel,OSP}^{peak}$ at SP2 and SP4 are similar. This may indicate that exhaust performance of detached XPT has weaker dependence on $dr_{Xpt}$, unlike attached XPT (fig 9), but further experiments are needed to confirm this.

Overall, we find that the improved heat flux mitigation capabilities of ADCs persist in the detached ELMy H-mode regime. Remarkably lower $q_{\parallel,OSP}^{peak}$ in detached H-mode ADCs, in addition to lower $J_{sat}^{peak}$ and $\Gamma_{ion}$, while maintaining high core confinement (as discussed earlier), make both ADCs desirable candidates for detached H-mode operation in future tokamaks. While we see the improved heat flux mitigation at the OSP/OSPs in ADCs, it is also important to investigate the effect of ADCs at the inner strike-point (ISP), in case the missing heat-flux isn't simply diverted to the ISP. This is addressed in the following section.

### *5.3 Heat fluxes at the ISP*

Figure 12 shows the radial profiles of the inter-ELM parallel heat fluxes at the ISP ($q_{\parallel,ISP}$) for the SN and the ADCs with baffles in the attached (black) and detached (color) phase. These profiles are taken from the same shots and time windows as in fig.11. In the attached phase, the peak $q_{\parallel,ISP}$ ($q_{\parallel,ISP}^{peak}$) is found to be over 2 times higher than at the OSP in the SN as well as the ADCs. $q_{\parallel,ISP}^{peak}$ in the ADCs remains lower than in the SN. The XPT exhibits the lowest

$q_{\parallel,ISP}^{peak}$, i.e., ~50% lower than in the SN's attached phase. Earlier detachment experiments in L-mode indicated that the ISP is harder to detach in TCV [15], which is also the case here. With $N_2$ seeding, the XD shows the highest (~80%) reduction in $q_{\parallel,ISP}^{peak}$, followed by the SN (~65%).

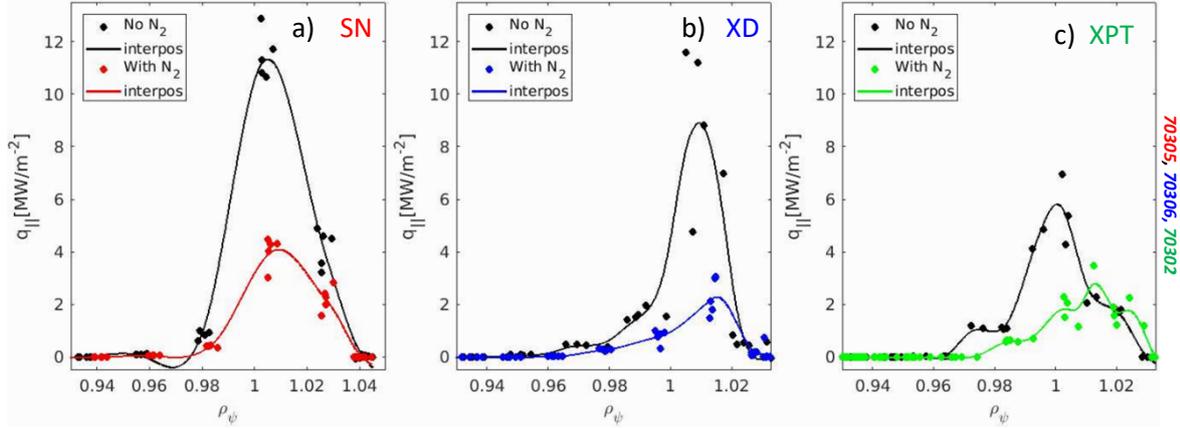

*Figure 12: Plots showing radial profile of $q_{\parallel,ISP}$ in baffled a) SN, b) XD, and c) XPT before (black) and after $N_2$ seeding (coloured) in $P_{NBI}$= 1.3 MW shots.*

The weakest effect of $N_2$ seeding is observed in the XPT with only a ~50% reduction in the $q_{\parallel,ISP}^{peak}$. Nonetheless, the $q_{\parallel,ISP}^{peak}$ in both ADCs remains lower than the SN, particularly in the XD. This evidence, again, reinforces the benefits of ADCs over SN.

It can be concluded from these results that the decrease in $q_{\parallel,OSP}$ for ADCs is not related to a redirection of power to the ISP. The effect of ADCs on different power loss channels in attached and detached H-mode will be discussed in the following section.

## 6   Effect of geometry on power distribution

The main power loss channels in these H-mode plasmas, apart from the previously discussed inter-ELM target heat fluxes, are impurity and hydrogenic radiation and heat pulses from ELMs. To understand the underlying mechanism leading to improved target inter-ELM heat flux mitigation in the ADCs, power loss through different channels is studied and discussed in this section.

### 6.1 Radiated power distribution in different geometries

Radiation emission profiles are studied using the newly upgraded RADCAM system in TCV [60]. Figure 13 plots the 2D emissivity profiles (filtered to remove ELM times) for detached H-mode for the SN and the ADCs with high power (1.3 MW). A strong radiation region localized near the X-point in all the geometries suggests we are in, or close to, X-point radiator regimes, like those observed in impurity seeding experiments in AUG [64]. Throughout, radiation is peaked around the X-point during the H-mode. However, unlike extreme X-point radiator regimes in AUG, ELMs are active in these discharges as noted earlier. These 2D profiles do not indicate that there is any remarkable influence of geometry on radiation. For a better comparison, the radiation from different regions in the vessel volume, namely core, divertor, and X-point (marked in fig. 13a) were calculated by integrating their emissivity. Figure 14 shows the radiated power from these regions together with that in the total vessel volume. The total radiated power in the ADCs and the SN are quite similar in both the attached and detached H-mode phases. In the attached phase, the total radiated power ($P_{rad}$) is ~ 0.6

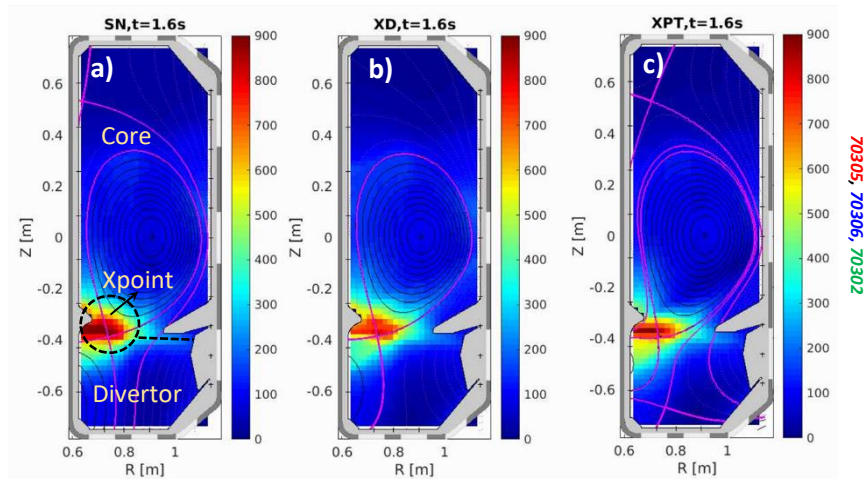

*Figure 13: Plot showing 2D emissivity profiles from bolometry in baffled a) SN, b) XD, and c) XPT in $P_{NBI}$ = 1.3 MW shots during the detached phase.*

MW in all geometries. After the N2 seeding, $P_{rad}$ increases to ~ 0.75-0.8 MW i.e ~ 85% of the total power coupled to the plasma in these geometries. The radiated power in the core and the X-point region follow a similar trend. Before the N2 seeding, the radiated power from the X-point and the divertor regions are similar, ~0.15 MW in ADCs and the SN. With N2 seeding, radiation from the divertor starts to decrease marginally whereas radiation from the X-point increases in all geometries. In the detached phase, the radiated power from the X-point is 40% higher than in the divertor region. However, no strong difference in radiated power nor its distribution is observed between the ADCs and the SN that could immediately explain the improved target heat flux mitigation in the ADCs. Similar results were also observed in the low power ADCs and SN.

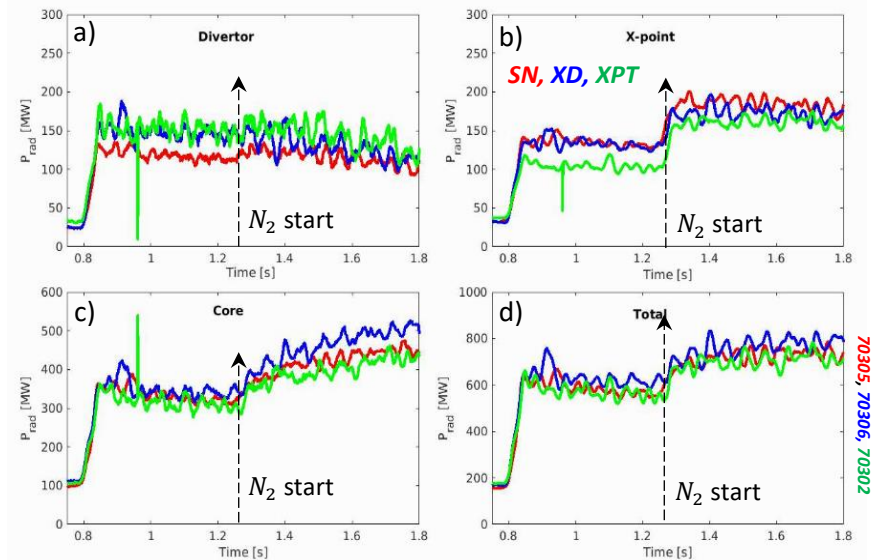

*Figure 14: Plots showing the time evolution of radiated power from a) divertor, b) X-point, c) core, and d) total vessel volume in SN (red), XD (blue) and XPT (green) in $P_{NBI}$ = 1.3 MW shots.*

The effect of baffles on radiation in these discharges is similar to that reported in Ref. [38].

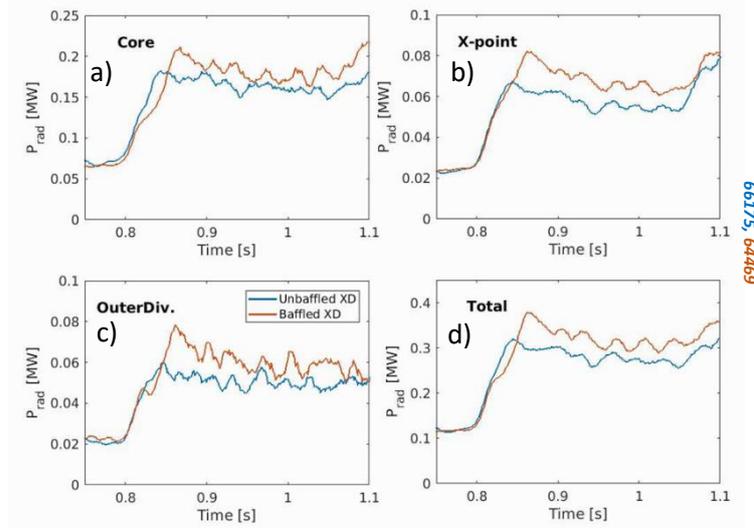

*Figure 15 : Plots showing the time evolution of radiated power from a) divertor, b) X-point, c) core, and d) total vessel volume in unbaffled (blue) and baffled (orange) XD in $P_{NBI}$= 0.75 MW shots.*

The baffled plasmas for all geometries exhibit higher radiated power as compared to the unbaffled plasma. Figure 15 shows the radiated power from the different regions of a baffled and an unbaffled XD plasma (0.75 MW). Increase in radiated power with baffles was observed in both ADCs and the SN. This is qualitatively consistent with the observed lower heat fluxes in the baffled scenarios.

The reason behind a reduced target heat flux in the ADCs compared to the SN could not be deduced from the radiated power analysis. An attempt to understand the power balance in the ADCs and the SN is made in the following section.

## 6.2 Power balance in SN and ADCs

The power balance in these H-mode plasmas is written as:
$P_{in} = P_{rad} + P_{target} + P_{ELM}$ where $P_{in}$ is the ohmic and NBI power delivered to the plasma after accounting for beam duct and orbit losses, calculated using the ASTRA code [65]. $P_{target}$ is the total inter-ELM power deposited on the divertor and is calculated by integrating the inter-ELM heat flux across all the strike points, assuming toroidal symmetry [51]. $P_{rad}$ is the total radiated power and $P_{ELM}$ is the power loss due to the ELMs. Figure 16 shows the time evolution of $P_{in}, P_{rad}, P_{target}$ and $P_{ELM}$ for the ADCs and the SN for the 1.3 MW baffled scenarios from fig 2. Both the ADCs and the SN have similar power distributions in terms of $P_{in}, P_{rad}$ and $P_{ELM}$ in the attached and detached H-mode. We find $P_{ELM}$ comparable to $P_{target}$ for all geometries. The effect of divertor geometry is evident only for $P_{target}$. This highlights also that the effect of divertor geometry on the peak inter-ELM heat fluxes at the SPs is more pronounced (as discussed in sec.5) than on the total power deposited at the SPs, particularly in the detached phase.

The power deposited at the divertor target is, clearly, only a small fraction i.e., ≤ 20% of the total radiated power. The total radiated power accounts for ~60% and 85% of the $P_{in}$ in the attached and detached phases, respectively. The absolute difference in $P_{target}$ between geometries is relatively small within the level of fluctuation in the total $P_{rad}$ data. Nevertheless, the XD geometry exhibits slightly higher radiated power throughout the attached and the detached phases. Similar trends were also observed in lower power plasmas. This small but consistent difference in radiated power could explain, at least in part, the reduced target heat

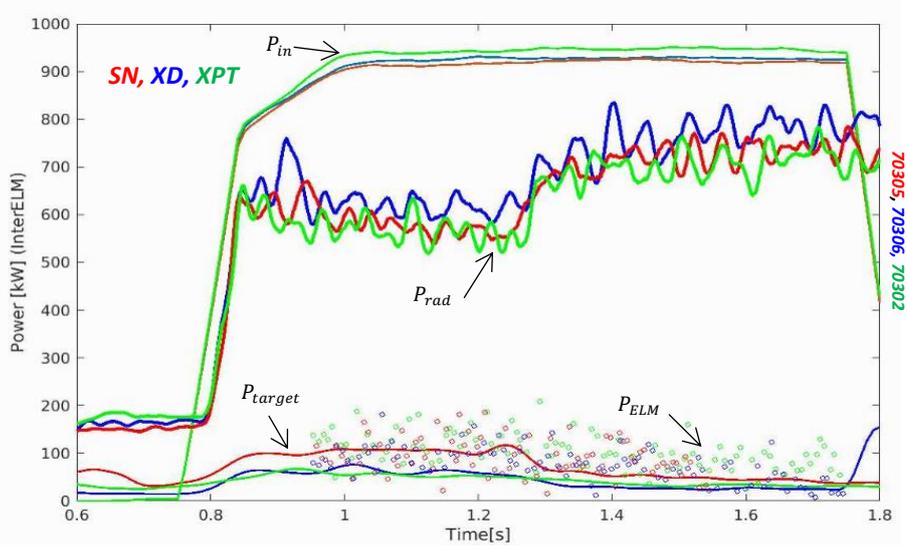

*Figure 16: Plot showing power distribution in baffled SN (red), XD (blue) and XPT (green) in $P_{NBI}$ = 1.3 MW shots.*

flux in the XD plasmas.

These results suggest that even in discharges with high radiated power fractions and radiation localised around the X-point, the divertor leg geometry still influences the target heat fluxes. The physics behind this heat flux reduction remain unclear. It might be due to the differences in radial transport in ADCs leading to the reduction in the peak heat fluxes. Power loss due to non-coronal processes, which are predicted to be enhanced in baffled XD, might also be responsible for this reduction [51]. Another possibility can be the difference in neutral losses, which can be enhanced in the XD [47].

## 7. Conclusions

High confinement detached H-mode ADCs, namely X-Divertor (XD) and X-Point Target (XPT), with improved heat and particle flux mitigation at the outer target, have been achieved with $N_2$ seeding at TCV. A detailed comparative study of many core and target parameters between the ADCs (XD and XPT) and the Single- null (SN) has been presented in this paper. The benefits of baffles, as reported in earlier experiments [38], are reproduced in all geometries. Compared to the baffled SN, a remarkable decrease in inter-ELM $q_{\|,OSP}^{peak}$ (>50%) and $J_{sat}^{peak}$ is achieved in ADCs for both attached and detached phases, whilst maintaining high core confinement $H_{98} \geq 1$. Figure 16 shows a simplified summary of the effect of adding baffles, modifying geometry and $N_2$ seeded detachment, respectively. Compared to an unbaffled, attached SN, ~95%-98% reduction in $q_{\|,OSP}^{peak}$ is achieved as a synergistic effect of adding baffles, modifying the divertor geometry and impurity seeded detachment. We have shown that the reduction in $q_{\|,OSP}^{peak}$ is not a redistribution of $q_\|$ between outer and inner strike points in the

ADCs, with the $q_{\parallel}^{peak}$ at the inner target also lower in the ADCs. The power distribution analysis indicates marginally higher radiated power in the XD, however, the physics behind

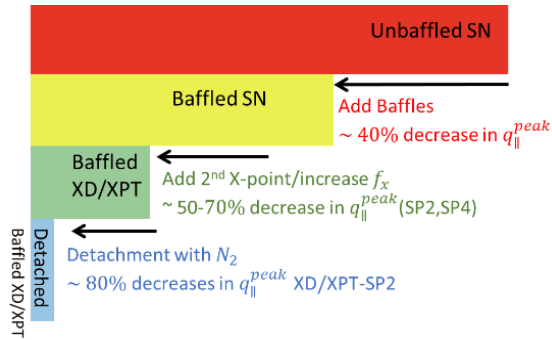

*Figure17: Schematic summarising the effect of adding baffle, modifying divertor geometry and impurity seeded detachment on $q_{\parallel,OSP}^{peak}$*

the reduced particle/heat flux to the target in ADCs remains unclear. These results exhibit a promising potential of XD and XPT in future reactors motivating continued investigation of these geometries.

In future, we plan to expand the ADCs in TCV even experiments to higher power scenarios with combined NBI and ECRH heating. The advantages of ADCs will be assessed in further operational regimes such as small ELM scenarios. Initial results from ADC experiments in such small-ELM regimes support the results in this paper. Experiments with increased poloidally and toroidally distributed impurity seeding uniformity (multiple gas injection points) in ADCs are also being developed, to seek optimisation of the impurity injection in TCV. Efforts are also continued to initiate numerical modelling of these ADCs in TCV, which is thought to be instrumental in understanding the underlying physics behind the improved exhaust performance of ADCs.

## 8. Acknowledgement

This work was supported in part by the Swiss National Science Foundation. This work has been carried out within the framework of the EUROfusion Consortium, funded by the European Union via the Euratom Research and Training Programme (Grant Agreement No 101052200 — EUROfusion). Views and opinions expressed are however those of the author(s) only and do not necessarily reflect those of the European Union or the European Commission. Neither the European Union nor the European Commission can be held responsible for them.